\documentclass[12pt]{iopart}

\usepackage{amsbsy}
\usepackage[english]{babel}
\usepackage{graphicx}
\usepackage{units}

\begin{document}

\title[Splitting of the Kondo resonance in anisotropic magnetic
impurities]{Splitting of the Kondo resonance in anisotropic magnetic
impurities on surfaces}

\author{R \v{Z}itko $^{1,2}$, R Peters $^1$ and Th Pruschke $^1$}
\address{$^1$ Institute for Theoretical Physics, University of G\"ottingen,
Friedrich-Hund-Platz 1, D-37077 G\"ottingen, Germany}
\address{$^2$ J. Stefan Institute, Jamova 39, SI-1000 Ljubljana, Slovenia}
\ead{rok.zitko@ijs.si}

\date{\today}

\begin{abstract}
Using the numerical renormalization group method, we study the splitting of
the Kondo resonance by a magnetic field applied in different directions in
the Kondo model for anisotropic magnetic impurities. Several types of 
magnetic anisotropy are considered: the XXZ exchange coupling anisotropy
$J_\perp \neq J_z$, the longitudinal magnetic anisotropy $DS_z^2$, and the
transverse magnetic anisotropy $E(S_x^2-S_y^2)$. In the spin-$1/2$ model
with the XXZ exchange coupling anisotropy we find very small direction
dependence in the magnitude of the splitting. In the spin-$3/2$ model with
the easy-plane ($D>0$) anisotropy, we observe very unequal magnitudes with
further differences between $x$ and $y$ directions in the presence of an
additional transverse anisotropy. A simple and rather intuitive
interpretation is that the splitting is larger in magnetically soft
directions. The magnitude of the splitting is directly related to the energy
differences between spin states and it is only weakly modified by some
multiplicative factor due to Kondo screening. The results for the $S=3/2$
model are in good agreement with recent scanning tunneling spectroscopy
studies of Co impurities adsorbed on CuN islands on Cu(100) surfaces [A. F.
Otte et al., Nature Physics {\bf 4}, 847 (2008)].
\end{abstract}


\maketitle

\newcommand{\vc}[1]{{\boldsymbol{#1}}}
\newcommand{\ket}[1]{|#1\rangle}
\newcommand{\bra}[1]{\langle #1|}
\newcommand{\braket}[1]{\langle #1 \rangle}
\renewcommand{\Im}{\mathrm{Im}}
\renewcommand{\Re}{\mathrm{Re}}
\newcommand{\dr}{\mathrm{d}}
\newcommand{\correl}[1]{\langle\langle #1 \rangle\rangle_\omega}
\newcommand{\TKO}{T_K^{(0)}}
\newcommand{\TKt}{T_K^{(2)}}

\newcommand{\figw}{7cm}
\newcommand{\dIdV}{\mathrm{d}I/\mathrm{d}V}

\bibliographystyle{unsrt}

\section{Introduction}

Transition metal atoms with partially filled $d$ shells tend to form local
magnetic moments. When such magnetic atoms are embedded in simple metallic
hosts or adsorbed on their surfaces, they induce various anomalies in the
low-temperature thermodynamic and dynamic properties (Kondo effect). The
most distinctive feature is the presence of a sharp resonance in the density
of states near the Fermi level which is known as the Abrikosov-Suhl
resonance or the Kondo resonance. This resonance cannot be interpreted in
the single-electron picture; it is of many-particle origin due to enhanced
exchange scattering of the low-energy conduction-band electrons which screen
the impurity moment. By applying an external magnetic field, the impurity
becomes spin polarized, the Kondo resonance splits and the Kondo effect is
suppressed when the Zeeman energy $g \mu_B B$ is increased beyond the
characteristic energy scale $k_B T_K$, where $T_K$ is the Kondo temperature.

Experimental attempts to directly observe the Kondo resonance in dilutely
doped metals using the photoemission spectroscopy are beset with
difficulties in measuring spectral features near the Fermi level, weak
signals, and limited energy resolution.
A more favourable approach is the transport spectroscopy of quantum dots with
non-zero total spin of the confined electrons \cite{goldhabergordon1998b,
cronenwett1998, kogan2004, quay2007}; these nanostructures can be considered
as artificial magnetic atoms with an additional benefit of being easily
tunable by changing potentials on the external electrodes. 
In these systems the Kondo effect manifests as a zero-bias anomaly in the
transport properties and the Kondo resonance can be probed by measuring the
differential conductance as a function of the bias voltage, although the
interpretation of these results is not straightforward since the quantum dot
is driven out of equilibrium \cite{meir1993}. The splitting of the Kondo
resonance was observed by applying a magnetic field \cite{meir1993,
goldhabergordon1998b, cronenwett1998}, but there remain open questions
regarding the magnitude of the splitting \cite{quay2007}.

Recently, real magnetic impurity atoms adsorbed on noble metal surfaces were
probed using scanning tunneling microscopes (STM) \cite{li1998,
madhavan1998}. The hybridization of the impurity with the substrate is
relatively strong in these systems (chemisorption) and the tip of the STM
acts essentially as a non-perturbing probe of the local density of states at
the adsorption site; the impurity is thus at equilibrium with the substrate
metal. Rather than a Lorentzian-like Kondo resonance, a Fano-resonance-like
feature is seen in the $\dIdV$ spectra, presumably because the electrons are
mostly tunneling into the $sp$-like adatom levels which hybridize strongly
with the substrate and extend the furthest out into the vacuum region, while
direct tunneling into the strongly localized inner $d$-shells is unlikely;
this interpretation is confirmed by the very weak dependence of the
tunneling spectra on the tip-surface separation. Strong hybridization
results in relatively high Kondo temperatures (tens of $\unit{K}$), thus
laboratory magnetic fields are not strong enough to induce an observable
Kondo resonance splitting (at most one could expect to observe a small
broadening of the Kondo resonance which might be within the energy
resolution of ultra-low-temperature STMs). It is, however, also possible to
adsorb magnetic atoms on ultra-thin ``isolating'' CuN islands on the Cu(100)
substrate, thereby reducing the Kondo temperature to the more appropriate
$\unit[1]{K}$ range. The splitting of the Kondo resonance in the magnetic
field can then be easily observed using fields in the $\unit{T}$ range
\cite{heinrich2004, otte2008}.
Furthermore, in these systems the zero-bias anomaly in the $\dIdV$ spectra
takes the simple form of a Lorentzian-like resonance, which makes the
comparison with theory more straightforward. In recent spin-excitation
spectroscopy measurements using an STM, the magnetic field was applied in
different directions in the Co/CuN/Cu(100) system, uncovering strongly
anisotropic Kondo resonance splitting with magnitudes differing by a factor
of two \cite{otte2008}. 
Prior studies have demonstrated that such magnetic adsorbates may be well
described by simple Kondo-like impurity models with magnetic anisotropy
terms \cite{hirjibehedin2007}. In this paper, we thus study spectral
functions of high-spin Kondo models with various sources of magnetic
anisotropy in an external magnetic field. In the case of the $S=3/2$ Kondo
model with easy-plane magnetic anisotropy, as relevant for the experiments
in the Co/CuN/Cu(100) system, we find results which agree very well with the
experiment.

\section{Model and method}

We consider the anisotropic Kondo impurity model \cite{anderson1970,
konik2002, schiller2008, costi1996akm, costi1998akm, romeike2006,
romeike2006b, wegewijs2007}
\begin{eqnarray}
H &  =
\sum_{\vc{k}\sigma} \epsilon_{\vc{k}} 
c^\dag_{\vc{k}\sigma} c_{\vc{k}\sigma}
+ J_z s_z S_z 
+ J_\perp \left( s_x S_x + s_y S_y \right) \\
& + D S_z^2 + E(S_x^2-S_y^2)
+ \sum_\alpha g_\alpha \mu_B B_\alpha S_\alpha.
\end{eqnarray}
Operators $c_{\vc{k}\sigma}$ describe conduction band electrons with
momentum $\vc{k}$, spin $\sigma \in \{ \uparrow, \downarrow \}$, and energy
$\epsilon_\vc{k}$, while $\vc{s}=\{s_x,s_y,s_z\}$ is the spin-density of
the conduction-band electrons at the impurity site.
Furthermore, $\vc{S}=\{S_x,S_y,S_z\}$ are the quantum mechanical impurity
spin-$S$ operators, $J_z$ and $J_\perp$ are the longitudinal and transverse
Kondo exchange coupling constants, $D$ is the longitudinal and $E$ the
transverse magnetic anisotropy. Finally, the $g_\alpha$ are the g-factors,
$\mu_B$ the Bohr magneton and $B_\alpha$ the magnetic field; index $\alpha
\in \{x,y,z\}$ denotes a direction in space. Note that we have assumed that
the magnetic anisotropy tensor and the $g$ tensor are diagonal in the same
frame; this is not necessarily always the case.

The Kondo resonance splitting in the $S=1/2$ Kondo model and in the Anderson
impurity model has been studied in the isotropic case by a number of
techniques: Bethe Ansatz \cite{moore2000, konik2001}, slave-boson
mean-field theory \cite{dong2001}, local-moment approach \cite{logan2001,
kim2007}, spin-dependent interpolative perturbative approximation
\cite{aligia2006interp}, and numerical renormalization group
\cite{costi2000, rosch2003, hewson2006, micklitz2007, wegewijs2007}. Some
analytical Fermi-liquid-theory results are known in the low-field limit
\cite{logan2001}, while the high-field limit is accessible by perturbation
theory. The non-trivial range is the cross-over regime, $g \mu_B B \sim k_B
T_K$.

We performed calculations using the numerical renormalization group (NRG)
method \cite{wilson1975, krishna1980a, bulla2008}. The results for
anisotropic impurities in the absence of the field were reported in
Ref.~\cite{aniso}; here we extend those studies to situations with a
magnetic field oriented in an arbitrary direction. Spectral functions are
computed using the density-matrix approach \cite{hofstetter2000} with
averaging over many interleaved discretization grids and using a narrow
broadening kernel in order to reduce errors due to over-broadening that
affect NRG results at high energies \cite{resolution}.

In the Kondo model, the Kondo resonance may be most conveniently observed in
the $T$-matrix spectral function \cite{koller2005, costi2000, zarand2004}.
The $T$-matrix is defined through $G=G^{(0)} + G^{(0)} T G^{(0)}$, where
$G^{(0)}$ and $G$ are the conduction-band electron propagators in the clean
system and in the system with impurity, respectively. The $T$-matrix
completely characterizes the scattering on the impurity and, in particular,
it contains information on both elastic and inelastic scattering
cross-sections \cite{zarand2004, borda2007}:
\begin{eqnarray}
\sigma_\mathrm{total}(\omega)/\sigma_0 &= \rho \pi^2 \left[ - \frac{1}{\pi}
\mathrm{Im} T(\omega) \right], \\
\sigma_\mathrm{el}(\omega)/\sigma_0 &= \rho^2 \pi^2 |T(\omega)|^2, \\
\sigma_\mathrm{inel}(\omega) &= \sigma_\mathrm{total}(\omega)
-\sigma_\mathrm{el}(\omega),
\end{eqnarray}
where $\sigma_0$ is the cross-section in the case of unitary scattering. We
also note that the $T$-matrix spectral function, $-(1/\pi) \mathrm{Im}
T(\omega)$, is the Kondo-model equivalent of the $d$-level spectral
function in the closely related Anderson impurity model. As a first
approximation, we may assume that the total scattering cross-section
is directly related (proportional) to the measured differential tunneling
current for magnetic adsorbates on decoupling layers.

\section{The splitting of the Kondo resonance}

We introduce $\Delta_\alpha=g_\alpha \mu_B B_\alpha$, i.e. the impurity
energy level spacing due to the magnetic field in the direction $\alpha$
(Zeeman splitting). The g-factors $g_\alpha$ cannot be easily determined in
experiments, but assuming that the orbital magnetism in the magnetic
adsorbate is completely quenched, we may as a first approximation assume an
isotropic $g$ tensor, $g_\alpha \equiv g_s \approx 2$. In theoretical
calculations, the actual values of $g_\alpha$ do not need to be known as
they enter the Hamiltonian only indirectly as energies $\Delta_\alpha$,
however they are important for the correct interpretation of experimental
results \cite{quay2007}. For Co/CuN/Cu(100), isotropic $g \approx 2.2$ was
established \cite{otte2008}.

We designate by $\delta_\alpha$ the position of the Kondo resonance, which
we will determine as a function of the Zeeman splitting $\Delta_\alpha$ for
a field applied in the direction $\alpha$. The $\delta$ vs. $\Delta$ curve
(in particular in the cross-over regime at intermediate fields $\Delta \sim
k_B T_K$) depends strongly on how the position $\delta$ is extracted from
the spectral functions. First we note that the Kondo resonance is not a
simple Lorentzian peak even in the zero-field limit, but it rather has
logarithmic tails \cite{resolution, bulla2000, dickens2001, glossop2002,
rosch2003}. Furthermore, the peak shape becomes increasingly asymmetric as
the magnetic field is established \cite{logan2001, rosch2003, hewson2006}.
As no unbiased peak fitting procedure can be introduced in general, it is
only meaningful to define the peak position as the energy of its maximum.
Further ambivalence arises here, since it is possible to extract the maximum
either from the spin-averaged spectral function (procedure A) or from
individual spin-dependent components (procedure B). The differences between
the results extracted by the two procedures are the most pronounced at the
onset of visible peak splitting in the spin-averaged spectral function, but
the two approaches become equivalent in the high-field limit. Since the peak
shapes themselves vary with $\Delta$, there is no generic way to relate the
results obtained using the two procedures. Experimentally one measures the
spin-averaged spectral function, while theoretically one is more interested
in the individual components, thus in this work we consider peak positions
as obtained by both procedures. Where needed, we will distinguish them by a
superscript, i.e. $\delta_\alpha^A$ vs. $\delta_\alpha^B$.

\subsection{Isotropic case}

For reference, let us consider first the known case of the isotropic $S=1/2$
Kondo model. In the limit of high magnetic fields, the splitting is expected
to be linear, i.e. $\delta=\Delta$ for $\Delta \gg k_B T_K$ (the distance
between the two peaks in the spin-averaged spectral function is thus exactly
twice the Zeeman energy); this is the usual Zeeman splitting for isotropic
free spins. In the low-field limit, the splitting is reduced by strong
correlations of the Kondo ground state and one expects to find linear
splitting, but with a different slope, i.e. $\delta=2/3 \Delta$ for $\Delta
\ll k_B T_K$ \cite{moore2000, logan2001, hewson2006}. The factor $2/3$ can
be derived using Fermi-liquid theory arguments \cite{logan2001}. The (slow)
cross-over between the two limiting regimes occurs around $\Delta \sim k_B
T_K$.

\begin{figure}[htbp]
\centering
\includegraphics[width=8cm,clip]{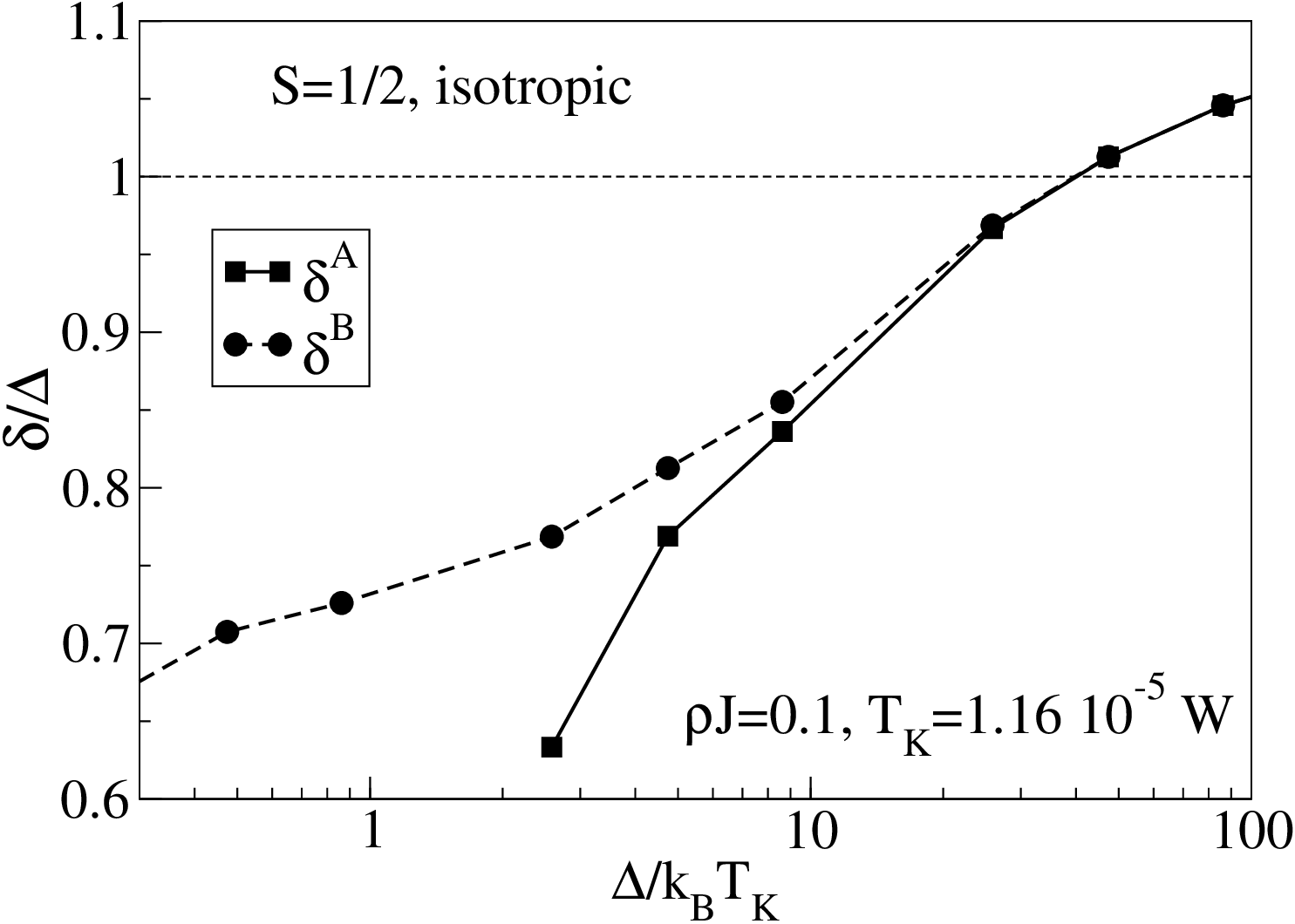}
\caption{
Kondo-resonance splitting in the isotropic $S=1/2$
Kondo model as a function of the magnetic field. We plot the position of the
Kondo peak, $\delta$, rescaled by the Zeeman energy, $\Delta=g\mu_B B$, as a
function of $\Delta/k_B T_K$. The peak position is extracted either from
spin-averaged spectral function ($\delta^A$) or from individual
spin components ($\delta^B$). Model parameters: $\rho J=0.1$, flat
band of width $2W$, so that $\rho=1/2W$.}
\label{figa}
\end{figure}

The Kondo-resonance splitting in the isotropic $S=1/2$ Kondo model with
$\rho J=0.1$, i.e. $T_K=1.16 \cdot 10^{-5} W$, is shown in Fig.~\ref{figa}.
The extraction of $\delta$ is difficult and error-prone in the small-field
limit (since the peak shift is much smaller than the peak width) and in the
large-field limit (due to NRG discretization artefacts), but it is
fortunately more reliable in the cross-over regime which is of main
interest. Nevertheless, deviations by a few percent from the universal
Kondo-scaling-limit results are expected for two reasons: 1) NRG artefacts
make it difficult to pinpoint the exact position of the peak maximum; 2) for
very large fields, $B$ becomes comparable to the scale of $J$ for the
present choice of model parameters and non-universal behaviour must occur. We
were thus not able to verify the approach to the expected limiting behaviours
$\delta=2/3 \Delta$ and $\delta=\Delta$. We note, in particular, that the
ratio $\delta/\Delta$ exceeds 1 in the large-field limit. For smaller $\rho
J=0.05$ with much reduced Kondo temperature $T_K=3\cdot 10^{-10}$, we do not
observe such behaviour for equivalent $\Delta/k_B T_K$, thus true universal
properties are only observed in the extreme Kondo limit.

\subsection{XXZ exchange anisotropy, $J_\perp \neq J_z$}

We studied the XXZ exchange anisotropic $S=1/2$ Kondo model with
$J_\perp=2J_z$ for $\rho J_z=0.1$. In the $S=1/2$ model, the exchange
anisotropy is an irrelevant perturbation and the system flows to the same
strong-coupling fixed point as in the isotropic Kondo model. Nevertheless,
while the fixed point itself is isotropic, the expansion around the fixed
point in terms of the irrelevant operators will be anisotropic with
expansion parameters which are functions of both $J_\perp$ and $J_z$, thus
dynamic response of the system is expected to exhibit some weak effects of
the anisotropy. We indeed find that the dynamic magnetic susceptibility is
somewhat larger in the transverse direction than in the longitudinal
direction at all frequencies. Anisotropic polarizability results in unequal
splitting of the Kondo resonance. We thus find that the splitting magnitude
is slightly larger for a field applied in the transverse direction. The
difference is largest for small fields (but still at most a few percent as
extracted by the procedure B) and it goes to zero in the large field limit
where the impurity again behaves as a free isotropic spin.

The XXZ exchange anisotropy is a relevant perturbation in the $S \geq 1$ 
Kondo models and it generates the $DS_z^2$ magnetic anisotropy term during
the renormalization process \cite{konik2002, schiller2008}. Therefore we
will not consider the effects of the XXZ exchange anisotropy in the $S \geq
1$ models here, since the qualitative behaviour of these models is similar to
that of models with bare $DS_z^2$ terms which are discussed in the
following.

\subsection{Longitudinal magnetic anisotropy, $DS_z^2$}

\begin{figure}[htbp]
\centering
\includegraphics[width=8cm,clip]{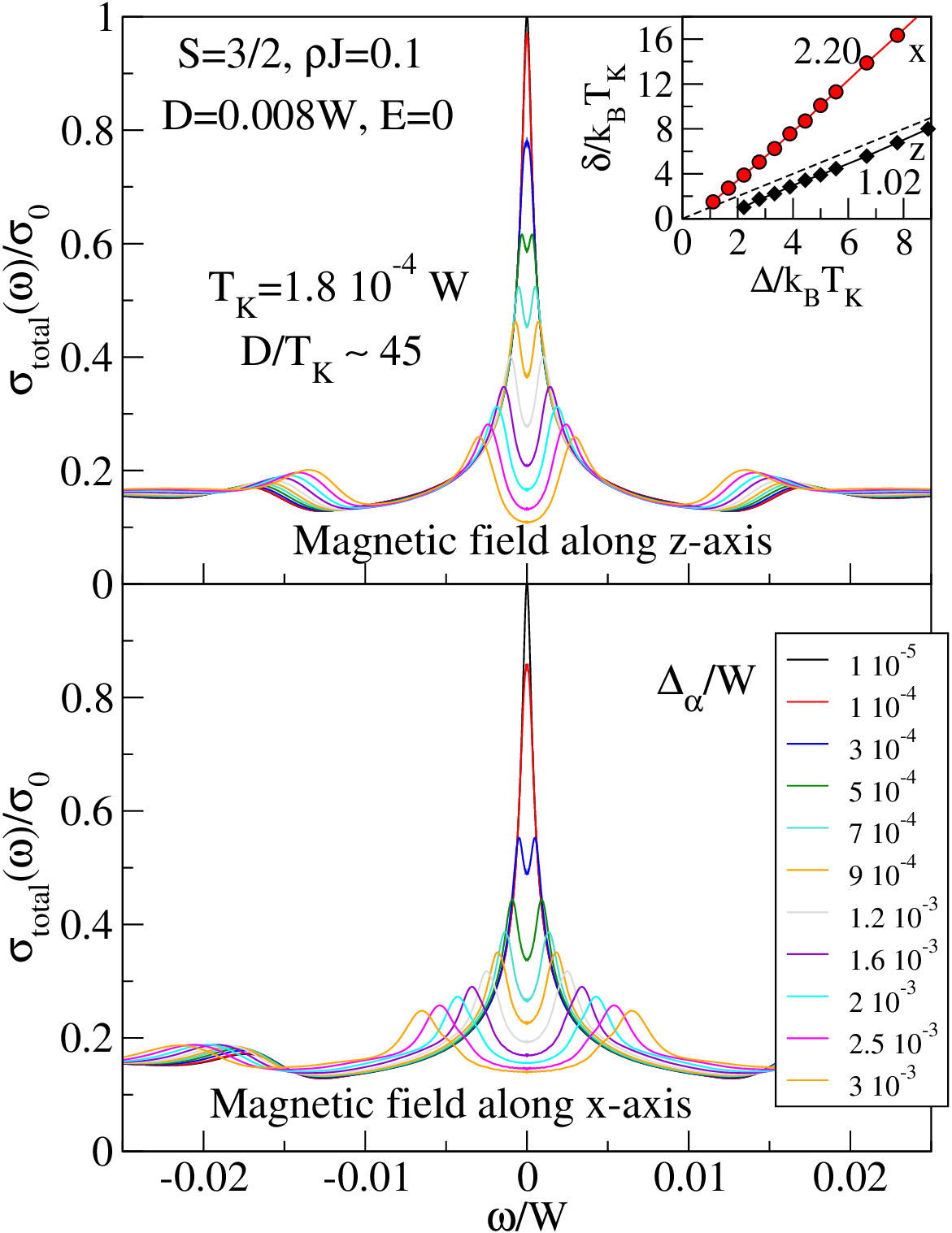}
\caption{%
Spectral functions for the $S=3/2$ Kondo model with easy-plane
anisotropy, $DS_z^2$ with $D>0$, for magnetic field applied in longitudinal
(upper frame) and transverse direction (lower frame). Inset: magnitude of the
Kondo peak splitting $\delta_\alpha^A$ vs. Zeeman energy $\Delta_\alpha$.}
\label{figc}
\end{figure}

We now study the effects of the longitudinal magnetic anisotropy term
$DS_z^2$ by considering the prototype $S=3/2$ Kondo model with easy-plane
anisotropy ($D>0$). This model undergoes effective $S=1/2$ Kondo effect if
$D>\TKO$, where $\TKO$ is the Kondo temperature for the isotropic $S=3/2$
model \cite{aniso}. In Fig.~\ref{figc} we plot spectral functions for a
range of magnetic fields. We find unequal splitting of the Kondo resonance
for fields applied in longitudinal and transverse direction with the ratio
of the splitting magnitudes (defined as the slopes of $\delta$ vs. $\Delta$
curves) being near two, see the inset in Fig.~\ref{figc}. The coefficient in
the $z$ direction is approximately 1 (as in the isotropic $S=1/2$ Kondo
model in the high-field limit), while the coefficient in the directions $x$
and $y$ is twice as large. There is also some offset: the extrapolated
$\delta$ vs. $\Delta$ curves do not pass through the origin. This is a
simple consequence of extracting $\delta$ by procedure A. We find that the
splitting coefficients are to a good approximation independent of $D$ as
long as $D \gg T_K^{(0)} \sim 10^{-5}W$ (2.2 and 1.1 for $D=0.001W$, 2.3 and
1.1 for $D=0.01W$, 2.2 and 1.0 for $D=0.1W$) and the ratio is thus quite
generically $\sim 2$. This is expected since the main role of the anisotropy
term $DS_z^2$ is to enforce a projection to the low-energy $|S_z|=1/2$
subspace. 

In the absence of any anisotropy and in zero magnetic field, Kondo screening
of the impurity spin by half a unit from $S=3/2$ to $S=1$ would occur on the
temperature scale $\TKO$ with a characteristically slow approach to the
unitary scattering limit (``underscreening''); the Kondo resonance is
cusp-like in this case \cite{posazhenikova2005, vzporedne2}. If $D < \TKO$,
the screening process stops abruptly on the temperature scale of $D$ and the
impurity spin becomes completely compensated \cite{aniso}. For $D > \TKO$,
however, the $|S_z|=3/2$ states freeze out on the scale of $D$, while the
low lying $|S_z|=1/2$ doublet behaves as an XXZ anisotropic spin-$1/2$ Kondo
model \cite{aniso}. For large $D$, the ratio of the effective exchange
constants is $J_\perp/J_z=2$; this stems from the fact that the $|S_z|=1/2$
submatrix of the operator $S_z$ is equal to $(1/2) \sigma_z$, while the
submatrices of the operators $S_x$ and $S_y$ are equal to $\sigma_x$ and
$\sigma_y$, i.e. twice the spin-$1/2$ operators \cite{aniso}. Furthermore,
and more importantly, in the $|S_z|=1/2$ subspace, the magnetic field
couples to the $x$ and $y$ components of the effective spin with twice the
usual strength $\Delta$. In other words, the g-factor of the effective
$S=1/2$ model in the transverse directions is twice as large as the
longitudinal g-factor. Having previously established that the XXZ exchange
anisotropy in the $S=1/2$ Kondo model leads to a difference of the magnitude
of splitting by only a few percent, the ratio of the splitting magnitude
around two in the $S=3/2$ model can be explained essentially by the
different g-factors.

\begin{figure}[htbp]
\centering
\includegraphics[width=8cm,clip]{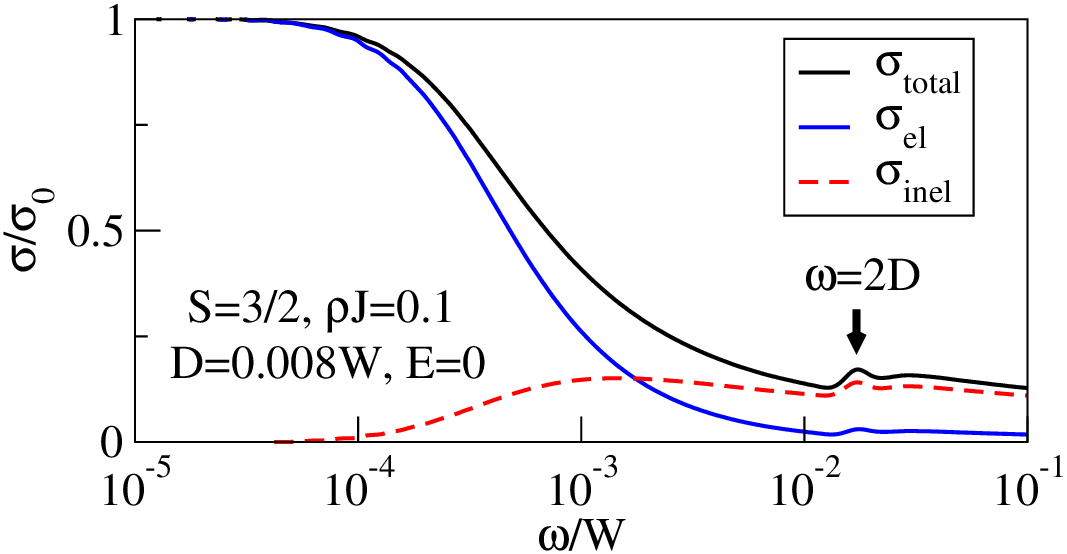}
\caption{%
Decomposition into elastic and inelastic scattering rates for the
anisotropic $S=3/2$ Kondo model.}
\label{figc1}
\end{figure}

There are additional features in the Kondo spectra at the excitation energy
$2D$ (at zero field). These peaks shift anisotropically as the magnetic
field is increased, similarly as the conductance steps in the experimental
$\dIdV$ plots \cite{otte2008}. They correspond to magnetic excitations from
the $|S_z|=1/2$ states to the $|S_z|=3/2$ states \cite{otte2008,
heinrich2004}. The weight in these spectral features is due mostly to
inelastic processes, see Fig.~\ref{figc1}. These magnetic excitation peaks
may be used to extract (experimentally) the g-factors if the magnetic field
strength is calibrated \cite{heinrich2004, otte2008}. 

\begin{figure}[htbp]
\centering
\includegraphics[width=8cm,clip]{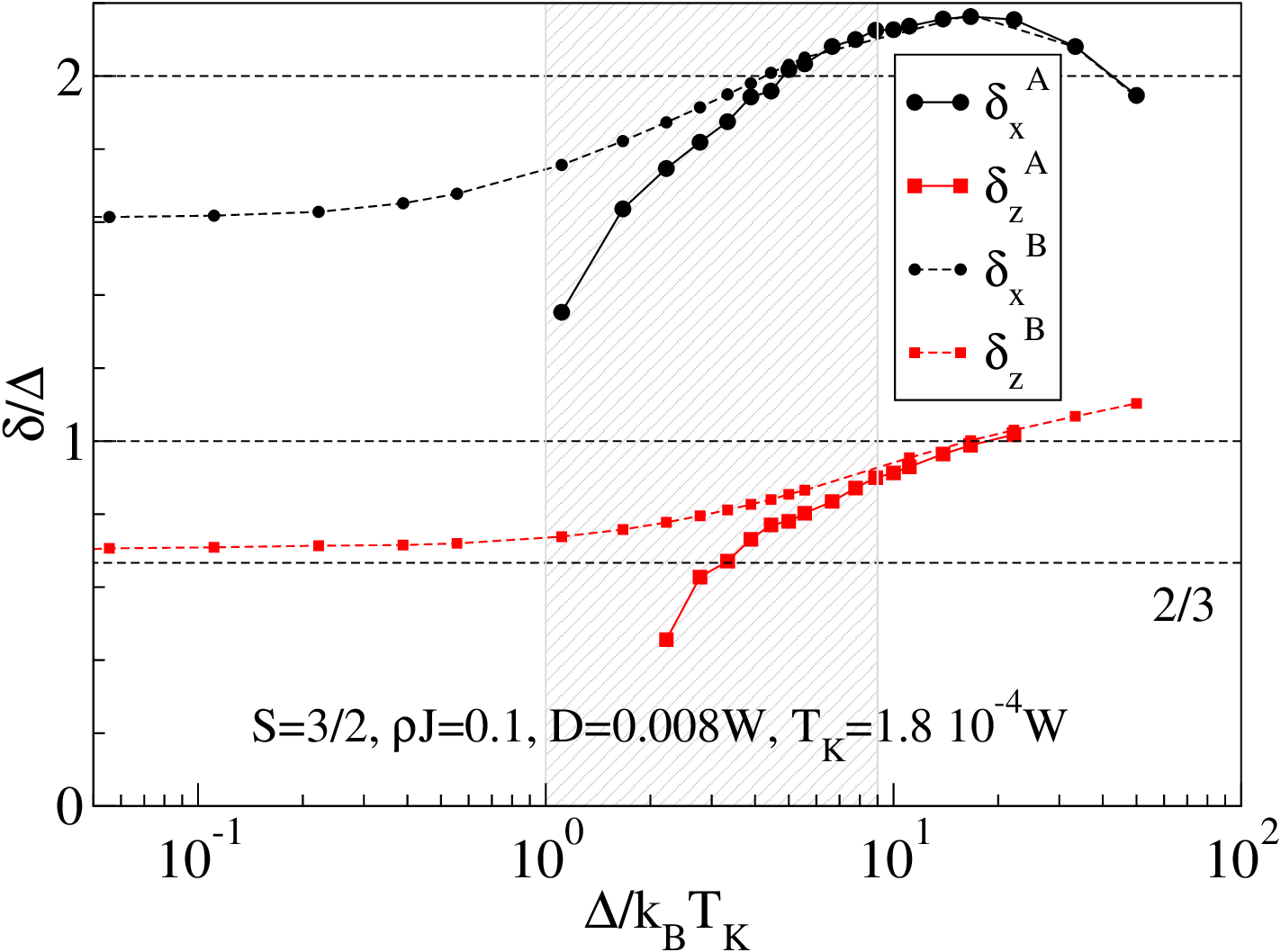}
\caption{%
Anisotropic Kondo-resonance splitting in the $S=3/2$ Kondo model
as a function of the magnetic field. The dashed region corresponds to the
``experimentally accessible range''; see also the inset in Fig.~\ref{figc}.}
\label{figc2}
\end{figure}

For $\Delta$ much smaller than $T_K$, the splitting can be extracted by
procedure B. This regime is not accessible experimentally using probes which
are not spin sensitive.
In Fig.~\ref{figc2} we plot the splitting ratio $\delta/\Delta$ as a
function of the rescaled Zeeman energy $\Delta/k_B T_K$ extracted using both
procedures (note that the splitting ratio $\delta/\Delta$ is not the same as
the slope of the ``experimental'' $\delta$ vs. $\Delta$ curve). We observe
the transition from the low-field $\Delta \ll T_K$ Fermi liquid behaviour to
the experimentally relevant intermediate-field behaviour, as well as
deviations for very strong fields where non-linear behaviour becomes manifest
for the magnetic field applied in the transverse direction. Remarkably, for
magnetic fields in the experimentally relevant range, the $\delta$ vs.
$\Delta$ curves are linear to an excellent approximation, even though the
curves in the low-field and high-field limits have more complex behaviour.
This is significant, since the results reported in
Ref.~\cite{otte2008} might find alternative interpretation as the
peaks appear to simply follow the eigenenergies of the decoupled spin
Hamiltonian, but we have shown that they are in agreement with the proposed
model (i.e. the splitting of a Kondo resonance). It may be noted that in the
calculations presented in Fig.~\ref{figc}, we had chosen a value of $D$ such
that the ratio $D/T_K$ is comparable to the experimental one, as determined
from the tunneling spectra shown in Fig. 2b in Ref.~\cite{otte2008}.

\subsection{Transverse anisotropy, $E(S_x^2-S_y^2)$}

\begin{figure}[htbp]
\centering
\includegraphics[width=8cm,clip]{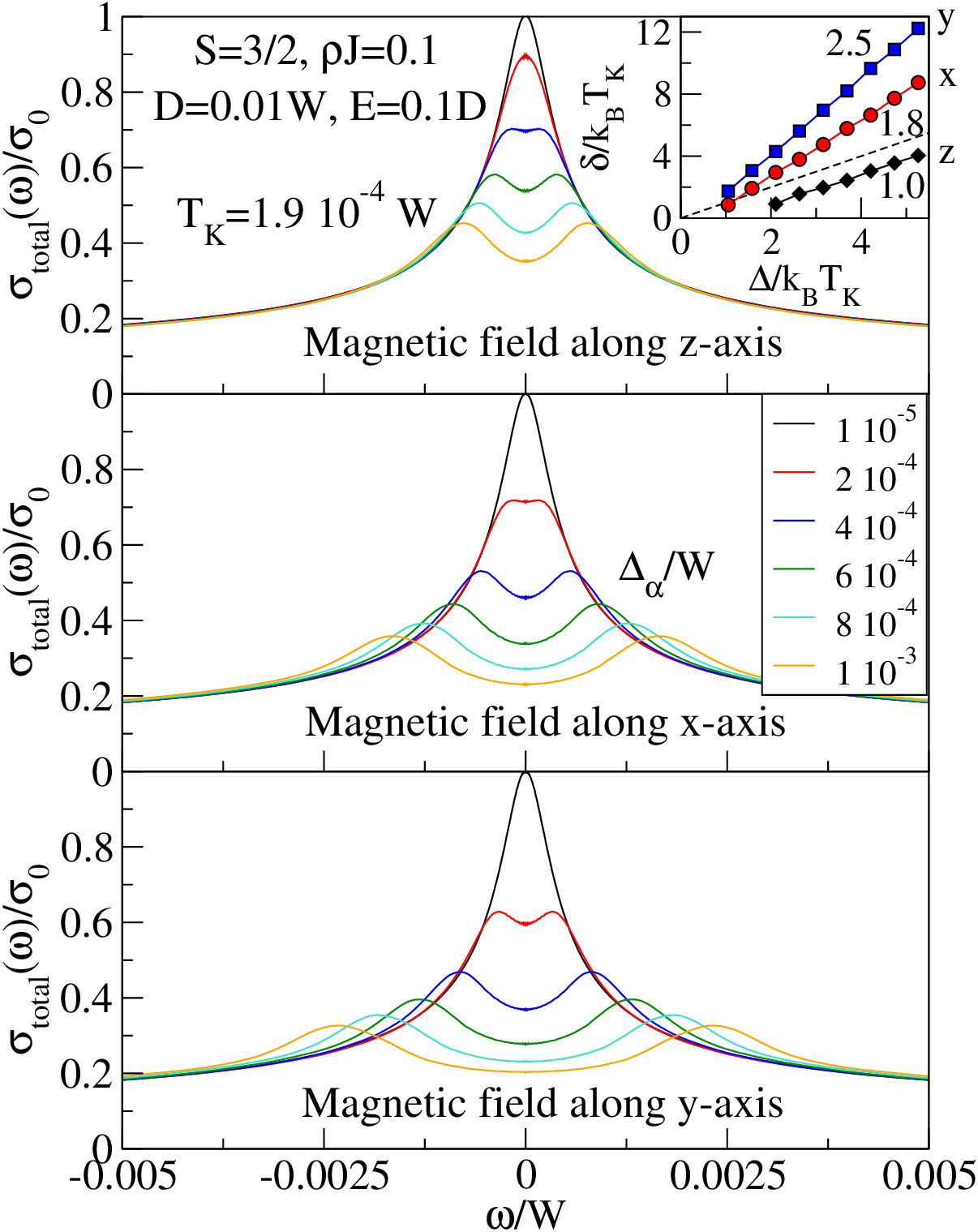}
\caption{%
Spectral function for the $S=3/2$ Kondo model with easy-plane 
anisotropy and weak transverse anisotropy.}
\label{figd}
\end{figure}

In the $S=3/2$ Kondo model with easy-plane anisotropy we now add additional
transverse anisotropy described by the term $E(S_x^2-S_y^2)$. We observe
that even a small transverse anisotropy $E$ leads to an appreciable
anisotropy in the directions $x$ and $y$, Fig.~\ref{figd}. In the
experimentally relevant range, the splitting curves are again linear to a
good approximation. The splitting magnitude increases to $2.5$ in the
``soft'' direction $y$ and it decreases to $1.8$ in the ``hard'' direction
$x$. The absence of such effects in the Co/CuN/Cu(100) system suggests that
the transverse anisotropy parameter $E$ is indeed very small.

In the presence of transverse anisotropy, the effective spin-$1/2$ degree of
freedom does not correspond to the $|S_z|=1/2$ states, but rather to two
degenerate linear combinations $\phi_{1,2}$ of the four $S_z$ states which
depend on the ratio $E/D$. In order to determine the effective g-factors in
this situation, one needs to project the spin-$3/2$ operators on the
$\phi_{1,2}$ subspace, which gives
\begin{eqnarray}
g^{\mathrm{eff}}_x &= \frac{1-3E/D}{\sqrt{1+3(E/D)^2}}+1, \\
g^{\mathrm{eff}}_y &= \frac{1+3E/D}{\sqrt{1+3(E/D)^2}}+1, \\
g^{\mathrm{eff}}_z &= \frac{2}{\sqrt{1+3(E/D)^2}}-1. 
\end{eqnarray}
For $E/D=0.1$, as relevant for the case presented in Fig.~\ref{figd}, we
find g-factors 1.69, 2.29, and 0.97, in fair agreement with the slopes of
the $\delta/\Delta$ curves shown in the inset.

\section{Conclusion}

On the example of the $S=3/2$ Kondo model with easy-plane magnetic
anisotropy which maps at low temperatures onto an anisotropic $S=1/2$ Kondo
model, we have shown that the effect of the external magnetic field may be
interpreted in terms of effective anisotropic g-factors which result
from different coupling of the magnetic field with the projected spin
operators in the $|S_z|=1/2$ subspace. The XXZ exchange coupling anisotropy,
also present in the effective $S=1/2$ model, plays a lesser role in this
respect, although this anisotropy is very important in that it strongly
increases the Kondo temperature of the spin-$1/2$ screening \cite{aniso}.

Generalizing these results to other high-spin models and to clusters of
coupled impurities which behave as effective anisotropic spin-$1/2$
impurities, we expect that the magnitude of the splitting of the Kondo peak
is given by the energy difference between the effective spin-$1/2$ states of
the decoupled impurity, multiplied by some prefactor of order 1 which is a
smooth function of $B/T_K$ and which takes into account the effect of the
Kondo screening. Since the prefactor varies very slowly (logarithmically)
with $B/T_K$, it may be taken to be a constant in the experimentally
relevant range of magnetic fields.

\ack
We acknowledge computer support by the Gesellschaft f\"ur wissenschaftliche
Datenverarbeitung (GWDG) in G\"ottingen and support by the German Science
Foundation through SFB 602 and project PR298/10.

\bibliography{aniso2}

\end{document}